\documentclass[11pt]{article}
\usepackage{DGfest,epsfig}
\usepackage{times}

\def\be{\begin{equation}}
\def\ee{\end{equation}}
\def\bea{\begin{eqnarray}}
\def\eea{\end{eqnarray}}
\def\lesim{${\lower 2pt\hbox{$\scriptstyle
<$}\atop\raise 4pt\hbox{$\scriptstyle\sim$}}$}
\def\gesim{${\lower 2pt\hbox{$\scriptstyle
>$}\atop\raise 4pt\hbox{$\scriptstyle\sim$}}$}

\begin{document}
\baselineskip 11.5pt
\title{ON THE INTERPLAY OF FERMIONS AND MONOPOLES IN\\ COMPACT QED$_3$}

\author{ SIMON HANDS$^a$, JOHN B. KOGUT$^b$ and BIAGIO LUCINI$^{a,c}$}

\address{$^a$~Department of Physics, University of Wales Swansea,
Singleton Park, Swansea SA2 8PP, U.K.}

\address{$^b$~Department of Energy, Division of High Energy Physics, 
1000 Independence Avenue SW, Washington
D.C. 20585-1290 \\
and\\
Department of Physics, University of Maryland,
82 Regents Drive, College Park, MD20742, U.S.A.} 

\address{$^c$~Institute for Theoretical Physics, ETH Z\"urich,
CH-8093 Z\"urich, Switzerland.}

\maketitle\abstracts{
The infra-red properties of three-dimensional abelian
lattice gauge theory are known to
be governed by a neutral plasma of magnetic monopole excitations. 
We address the fate of these monopoles in the presence of light dynamical
fermions, using a lattice formulation of compact QED$_3$ with $N_f=4$
fermion flavors supplemented
by a four-fermi contact term permitting numerical Monte Carlo
simulations in the chiral limit. 
Our data hint at a restoration of chiral symmetry above a
critical value of the (inverse) coupling $\beta$. By
performing simulations in a sector of non-vanishing magnetic charge, we are
able to study the response of the theory to an external magnetic test charge.
Our results suggest that the monopole plasma persists even once chiral symmetry
is restored, and hence survives the continuum limit.
}

\section{Introduction}

Adriano's outstanding interest in recent years has been an
understanding of color confinement in QCD.\cite{Ad1} 
His particular sense of beauty in physics has brought to
the field the idea that confinement can be understood in terms of a
symmetry.\cite{Ad2} The role of magnetic monopoles can then be exposed by
constructing a disorder parameter that tests the magnetic properties
of the vacuum.\cite{Ad3} Many of the concepts can be exported to
other theories where dual excitations are conjectured to play a
role in determining the phase structure of the system; 
compact abelian lattice gauge theory in 3 dimensions is one such theory.
This model, which we will refer to as ``quenched compact QED$_3$'', 
has been understood
for many years as a field theory example where interesting IR
effects such as linear charge confinement and generation of a mass gap 
arise as an effect of
monopoles~\cite{Polyakov,BMK,Stack} 
(which in $3d$ are not particles but
{\em instantons}). As the continuum limit is approached the system remains in 
the confining phase found at strong coupling, with
the vacuum thought of as a dilute charge-neutral plasma of monopoles $m$
and
antimonopoles $\bar m$ 
whose effect is to Debye-screen the Coulomb potential of a test
charge. The photon mass is proportional to $\exp(-C/e^2)$, $e$ being
the electric charge appearing in the relation between gauge potential and link
variables $U_{\mu x}=\exp(iaeA_{\mu x})$, where $a$ is the lattice spacing
and $C(a)$ related to the free energy of an
isolated Dirac monopole. Although this scale is $O(a^{-1})$ implying that the
monopole has a core extending over a few lattice spacings where UV artifacts
are significant, following Polyakov's original treatment~\cite{Polyakov}
the model can be viewed
as an effective
description of the Higgs phase of a $3d$ Georgi-Glashow model in which
semi-classical solutions exist for $m$ and $\bar m$, and hence a continuum limit
identified as $\beta\equiv1/e^2a\to\infty$.

More recently people have begun to ask if this situation persists
if light electrically-charged fermions are present, 
in which case the model is now ``compact
QED$_3$''. A simple way of seeing why this might be expected to have a dramatic
effect is to observe that as a result of the Dirac quantisation
condition the combination $eg$, where $g$ is the magnetic charge of the
monopole, is a renormalisation group invariant,\cite{Calucci} 
implying $g_R>g$. Hence, virtual
electron anti-electron pairs anti-screen the $m\bar m$ interaction. More
detailed calculations~\cite{Marston,Kleinert} suggest that $V_{m\bar m}(x)$
is modified from the $3d$ Coulomb form $1/x$ to $\ln x$, implying that a
deconfined phase with monopoles only present within tightly-bound $m\bar m$ 
molecules, and therefore unable to influence the IR physics, 
may exist for sufficiently large $\beta$. Other workers, however, have argued
that interaction among magnetic dipole pairs restores the Coulomb potential at
large $x$, and that confinement continues to survive the continuum 
limit.\cite{Herbut}

A related question is whether there exists a chiral symmetry restoring
transition at some finite $\beta_c$ (like confinement, spontaneous
chiral symmetry breaking leading to a mass gap for fermion excitations is known
to happen at strong coupling~\cite{Seiler}).
In {\em non-compact\/} QED$_3$ in which the Wilson action is replaced by 
a simple $F_{\mu\nu}^2$ term which is a non-periodic function of the plaquette,
so that monopoles are suppressed and play no role in the continuum limit,
this issue is believed to depend sensitively 
on the number of fermion species $N_f$. Chiral symmetry breaking is supposed 
to persist in the continuum limit for $N_f<N_{fc}$, where the critical value
$N_{fc}$ has been estimated as anything between ${3\over2}$
and $\infty$,\cite{Appelquist,Pennington}
with recent estimates based on truncated Schwinger-Dyson equations yielding
$N_{fc}\approx4$.\cite{Maris} Lattice estimates have proved
inconclusive due to the enormous separation of scales predicted by the SD
approach (eg. the combination $\langle\bar\psi\psi\rangle/e^2\sim
O(10^{-4}-10^{-5})$): so far studies on lattices up to $80^3$ suggest
$N_{fc}>1$.\cite{NCQED3} 

It is intriguing that in QED$_3$ 
the two classic non-perturbative phenomena are
accounted for by very different theoretical approaches; 
confinement is attributed to instanton effects whereas chiral symmetry
breaking
is accounted for by self-consistent solution of the SD equations.
An interesting issue raised by these considerations is 
whether compact and non-compact models
have the same continuum limit, which seems probable if indeed monopoles are
irrelevant for IR physics.
Are any other continuum scenarios
possible beyond a confining gapped theory and a deconfined chirally symmetric
one? How in fact is ``confinement'' characterised in the presence of dynamical
electric charges? How does everything depend on the parameter $N_f$? These
questions may have important 
potential applications in condensed matter 
physics,\cite{Marston,Kleinert,Herbut} and
have stimulated recent numerical work.\cite{Fiore,Ichinose}

\section{The Lattice Model}

In this paper we will present the results of numerical simulations of a variant 
of compact lattice QED$_3$ in which an extra four-fermion contact term has been
added to the action, which in terms of real-valued link potentials
$\theta_{\mu x}$ and $N_f/2$-component 
staggered fermion fields $\psi_x,\bar\psi_x$ reads
\be
S=\sum_{xy}\bar\psi_x(D_{xy}+M_{xy})\psi_y
+{\beta_s\over2}\sum_x\sigma^2_{\tilde x}
+\beta\sum_{x,\mu<\nu}(1-\cos\Theta_{\mu\nu x})
\label{eq:action}
\ee
where
$D_{xy}={1\over2}\sum_\mu\eta_{\mu x}(e^{i\theta_{\mu x}}\delta_{y,x+\hat\mu}
-e^{-i\theta_{\mu x}}\delta_{y,x-\hat\mu})$, $M_{xy}=({\cal
M}+{1\over8}\sum_{<\tilde x,x>}\sigma_{\tilde x})\delta_{xy}$,
and $\Theta_{\mu\nu x}=\Delta^+_\mu\theta_{\nu x}-\Delta^+_\nu\theta_{\mu x}$.
Here $\eta_{\mu x}$ are Kawamoto-Smit phases, 
$\sigma$ a real-valued scalar auxiliary field defined on the
dual lattice sites $\tilde x$, and $<\tilde x,x>$ denotes the set
of 8 dual sites neighbouring $x$. Gaussian integration over $\sigma$ yields
an attractive 
four-fermion interaction of the form $-G(\bar\psi\psi)^2$, with the coupling
$G\propto1/\surd\beta_s$. Pure compact lattice QED$_3$ is recovered in the limit
$\beta_s\to\infty$.

The four-fermi term has been introduced in studies of $4d$ models~\cite{KKL}
to enable simulations in the massless limit ${\cal M}=0$. Although 
in the limit $\beta\to\infty$ it induces
chiral symmetry breaking in its own right 
for some $\beta_s<\beta_{sc}$, 
in $4d$ the extra
interaction is irrelevant, and should leave the continuum limit
unaltered as $\beta_s$ is made large.  In $3d$ the situation is less clear: 
$\beta_{sc}\approx0.25N_f$ defines a UV-stable renormalisation group fixed
point,\cite{HKK} so that while we may suppose 
that as $\beta_s/\beta$ is increased 
any non-perturbative behaviour such as chiral symmetry breaking may be
attributed to gauge dynamics such as monopoles, it remains to 
be checked that this
behaviour is not simply associated with the fixed point of the $3d$ Gross-Neveu
model.
Note also that the four-fermi term reduces the chiral symmetry 
group from U(1) to Z$_2$.

Magnetic monopoles in the lattice model (\ref{eq:action})
are identified {\it\`a la} DeGrand-Toussaint:\cite{DeGT} for every plaquette the
Dirac string content is identified via
\be
\Theta_{\mu\nu x}=\overline\Theta_{\mu\nu x}+2\pi s_{\mu\nu x}
\ee
where $\overline\Theta\in(-\pi,\pi]$ and $s$ is integer. Gauge invariant 
integer magnetic charges on the
dual sites are then 
given by
\be
\tilde m_{\tilde x}=\epsilon_{\mu\nu\lambda}\Delta^+_\mu s_{\nu\lambda x}.
\ee
Since on a 3-torus the total magnetic charge $\sum_x\tilde m_{\tilde x}\equiv0$,
we find it useful to define the average magnetic charge magnitude
per site $m=V^{-1}\sum_x\vert\tilde m_{\tilde x}\vert$ as a measure of monopole
activity.

\begin{figure}
\begin{center}
\psfig{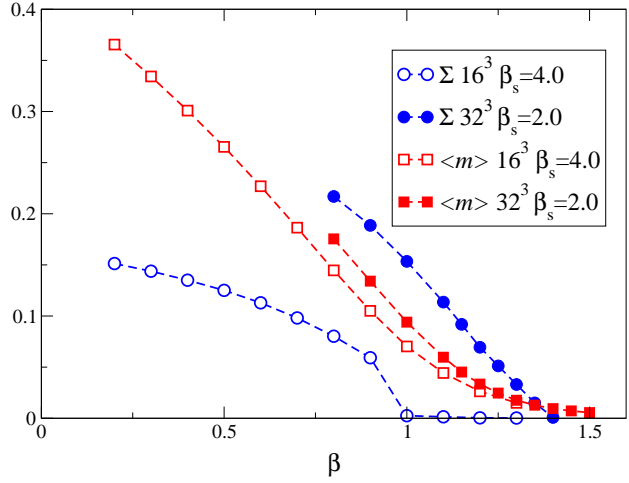}
\caption{Scalar condensate $\Sigma$ and magnetic charge density $\langle
m\rangle$ vs. $\beta$ for two values of $\beta_s$
\label{fig:overview}}
\end{center}
\end{figure}
We have performed simulations with $N_f=4$, ${\cal M}=0$, using a hybrid
molecular dynamics algorithm.
Fig.~\ref{fig:overview} sets the scene. We show both
$\Sigma\equiv\langle\sigma\rangle=\beta_s^{-1}\langle\bar\psi\psi\rangle$, 
and $\langle m\rangle$ as functions of $\beta$, 
for two four-fermi couplings $\beta_s=2.0$ and $\beta_s=4.0$.
Broadly speaking we see that chiral symmetry breaking is approximately restored
at large $\beta$, the transition occuring at a coupling $\beta_c(\beta_s)$
which grows stronger as $\beta_s$ is increased. 
There are corresponding peaks in the scalar susceptibility
at the same locations (see Fig.~\ref{fig:deltam_chi} below), 
suggesting that a true phase transition occurs.
There is not yet enough data on different volumes to make a definitive
statement about the nature or order of this transition; since it has proved so
difficult to identify $N_{fc}$ from simulations on finite
systems,\cite{NCQED3} we should not exclude the possibility of a first order
transition to a high-$\beta$ phase where chiral symmetry is still
very slightly broken. 
What we can say, though, is that the fermions in the weakly
coupled phase $\beta>\beta_c$ are much lighter.
We also see that $\langle m\rangle$ decreases monotonically with $\beta$, 
with both $\langle\bar\psi\psi\rangle$ and $\langle m\rangle$ becoming
``small'' at $\beta\approx1.5$ for $\beta_s=2.0$. This correlation between
chiral condensate and magnetic charge density was first observed by Fiebig and
Woloshyn,\cite{Fiebig} and more recently by Fiore {\it et al}.\cite{Fiore}
The correlation looks less convincing, however, for $\beta_s=4.0$, since
changing $\beta_s$ has relatively little effect on $\langle m\rangle$.

\section{A Fresh Approach}

As Adriano once memorably reminded us, ascribing properties such as confinement
to a system because of the presence of monopoles is a little like
claiming a metal is superconducting simply because it contains electrons. A more
sophisticated characterisation of the vacuum is needed. Unfortunately, here
it is impossible to construct a disorder parameter signalling a broken
symmetry associated with monopole condensation, as in $4d$ abelian
lattice gauge theory,\cite{disorder} because as mentioned above in $3d$
monopoles are not particles. Instead, we will follow an approach introduced
in studies of $3d$ non-abelian gauge theories,\cite{Hart} and look at  
the difference in free energy and its derivatives, and in particular their
dependence on lattice volume, when a magnetic test charge is
introduced. Physically this is akin to probing the
properties of the resulting magnetic field in the presence of both dynamical
monopoles and dynamical electrons: 
is it screened, indicating persistence of confinement,
or can its effects operate over arbitrarily large distances?

First let us review the construction, beginning with
the quenched case. The free energy
of the system is implicitly given by
\be
Z = e^{- \beta F} = \int {\cal D} \theta _{\mu} e ^{- S},
\ee
with as usual
\be
\label{u1action}
S = \beta \sum_{x, \nu > \mu} \left(1 - \cos \Theta_{\mu \nu x} \right).
\ee
Assuming that the action in the magnetic charge $q \ne 0$ sector has
the same form, we can rewrite
the same expressions for a system in the presence of an external monopole:
\be
Z_M = e^{- \beta F_M} = \int {\cal D} \theta _{\mu} e ^{- S_M},
\ee
\be
\label{u1mono}
S_M = \beta \sum_{x, \nu > \mu} \left (1 - \cos
\tilde{\Theta}_{\mu \nu x} \right).
\ee
with $\tilde{\Theta}_{\mu \nu}$ defined below.
The difference in free energy we are looking for is then formally
\be
\label{deltaf}
\Delta F = F_M - F = \frac{1}{\beta} \log \frac{Z}{Z_M}.
\ee

Because of the difficulty in measuring partition functions in lattice
simulations, we define~\cite{Luigi}
\be
\label{rho}
\rho = \frac{\partial}{\partial \beta} \left( \beta \Delta F \right) =
\langle S_M \rangle_{S_M} - \langle S \rangle_{S}.
\ee
The subscript indicates the action with respect to which the average
is taken. To determine $\langle S_M \rangle_{S_M}$
an independent simulation using the action $S_M$ instead of $S$ is needed,
implying a duplication of computational effort.
By integrating $\rho$ we could in principle determine $\Delta F$. However, 
unless we are interested in its precise value,  the general features of
the physics can be extracted directly from $\rho$: eg. we might expect it to
display a sharp peak in the critical region if monopoles have something to
do with any phase transition. We can also monitor other observables, such as
$\Delta m\equiv\langle m\rangle_{S_M}-\langle m\rangle_S$,
$\Delta\langle\bar\psi\psi\rangle$, etc.

For the form of the action~(\ref{u1mono}), a naive proposal is to use the
expression
\be
\tilde{\Theta}_{\mu\nu x} = {\Theta}_{\mu\nu x} + B_{\mu\nu x}, \qquad
\qquad \mu,\nu = 1,2,3 \ ,
\ee
where
$B_{\mu\nu x} = \Delta^+_\mu b_{\nu x} -
\Delta^+_\nu b_{\mu x}$
is the plaquette built with the link field
\be
\vec{b}(\vec{r}) = \frac{q}{2} \frac{\vec{r} \times  \hat{n}}{r(r -
\vec{r}\cdot  \hat{n})} \ ,
\ee
which is the vector potential of a Dirac monopole located at $\vec r=\vec0$
carrying $q$ units of
magnetic charge with string
along $\hat{n}$. 
The action in this form would have nice properties, like the correct continuum
limit and invariance under gauge transformations on the field $\vec{b}$.
However, if we adopt this form, we cannot use periodic boundary conditions.
In fact, it is easy to see that a simple link shift
$\theta_{\mu x}\mapsto\theta_{\mu x} + b_{\mu x}$
will reabsorb $\vec{b}$ into the partition function, so 
$Z_M = Z$ and no information on $\Delta F$ can be extracted. 

Our proposal is to keep standard boundary
conditions for all dynamical fields,
but to use free boundary conditions on the magnetic
field. In other words, the lattice topology seen by $\vec{b}$ is
not that of an hypertorus, but that of a standard cube. This ensures a
net magnetic flux exiting the box, and hence by Gauss's law a non-zero
magnetic charge. This ``mixed topology'' requires extra care for the
plaquettes on the bottom face (i.e. with a coordinate equal to 0) of
the lattice: since we have to take account also of the contribution
of the external field on the top face, these plaquettes enter the
action twice, with different $B_{\mu\nu}$.
The action then reads
\begin{eqnarray} 
\label{extaction}
S_M / \beta &=& \sum_{P \in {\cal B}} \left(1 -
\frac{1}{2}\cos (\Theta_{\mu \nu} + B_{\mu \nu}^{up}) -
\frac{1}{2}\cos (\Theta_{\mu \nu} + B_{\mu \nu}^{down})
\right) \label{eq:maction}\\
\nonumber
&+& \sum_{P \notin {\cal B}} \left(1 -
\cos \left(\Theta_{\mu \nu} + B_{\mu \nu} \right) \right) \ ,
\end{eqnarray}
where $P$ is the generic plaquette, ${\cal B}$ is the boundary of the
lattice and the superscripts $up$ and $down$ refer
respectively to the top and bottom faces of the
lattice.

When fermions are introduced, 
the interaction with the the external field entails
the following replacements in $D_{xy}$:
\be
e^{i \theta_{\mu x}} \to e^{i \left( \theta_{\mu x} + b_{\mu x}\right)} 
= e^{i \theta_{\mu x}}
\zeta_{\mu x}.
\label{eq:fmaction}
\ee
With ${\cal B^\prime}$ the boundary
without the edges, ${\cal E}$ the edges and ${\cal M}$ the bulk
without the boundary, and denoting a generic link
by $\ell \equiv (x,\mu)$ 
we have:
\be
\zeta_\ell = \left\{
\begin{array}{l}
 e^{i b_{\mu x}^{up}} \qquad \ell \in {\cal M}\\
\frac{1}{2} \left(e^{i b_{\mu x}^{up}} + e^{i b_{\mu x}^{down}} \right)
\qquad \ell \in {\cal B^\prime}\\
\frac{1}{4} \left(e^{i b_{\mu x}^{up1}} + e^{i b_{\mu x}^{up2}} +
e^{i b_{\mu x}^{up12}} + e^{i b_{\mu x}^{down}} \right)
\qquad \ell \in {\cal E}
\end{array}
\right.
\ee
A link in the edge must be identified with three other parallel links, here
called $up1$, $up2$ and $up12$. 
Note that the $\zeta$ couplings on the boundary are not unitary.

\section{The Quenched Theory}
\begin{figure}
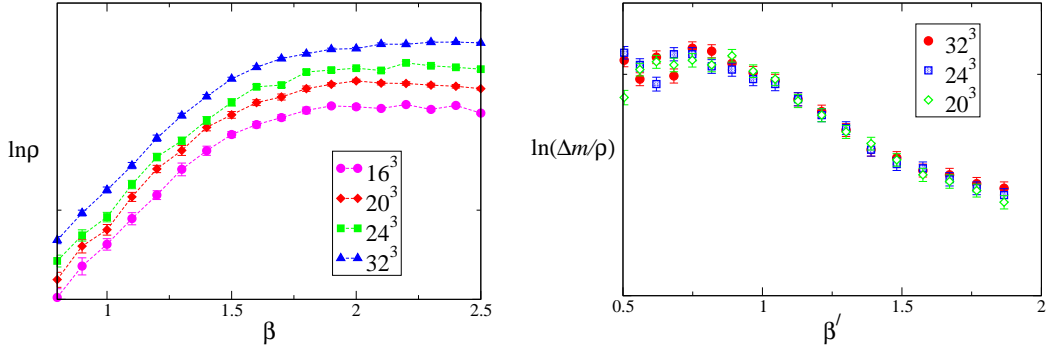

\psfig{figure=ln_rho.eps,height=1.8in}\hskip0.5cm
\psfig{figure=V_scaling.eps,height=1.8in}
\caption{Quenched results: 
(left) $\ln\rho$ and (right) $\ln(\Delta m/\rho)$ vs. $\beta$ for 
various lattice sizes.
\label{fig:quenched}}
\end{figure}
We first illustrate these ideas in the quenched model, where we ``know'' the
answer. The left panel of
Fig.~\ref{fig:quenched} presents $\ln\rho(\beta)$ 
for lattices ranging from
$16^3$ up to $32^3$, and shows a crossover from strong to weak coupling
behaviour at $\beta\approx1.5$, whose location is volume independent. The
behaviour at weak coupling can be accounted for by Polyakov's model of the
QED$_3$ vacuum as a dilute neutral monopole plasma. The
screened Coulomb potential is
\be
V(r)\propto{e^{-Mr}\over r}
\ee
with~\cite{Polyakov}
\be
M^2=4\pi^2\beta\exp(-\beta C)\equiv4\pi^2\beta\xi
\ee
where $\xi$ is the monopole fugacity.
We can use this effective potential to estimate the excess action in the 
presence of a test magnetic charge. The Lagrangian density ${\cal L}\propto
H^2(r)$, with $\vec H\propto-\vec\nabla V$. Therefore
\be
\rho=\int d^3r{\cal L}(r)\propto4\pi\int_a^L dr
e^{-2Mr}\left[M+{1\over r}\right]^2.
\label{eq:polyrho}
\ee
The UV contribution dominates the integral; since $Ma\sim\exp(-C\beta)$ we
deduce that at weak coupling 
\be
\rho\sim {{4\pi}\over a} e^{-2Ma}\propto\exp(-4\pi\surd\beta e^{-C\beta/2}),
\ee
and thus that $\rho$ depends only weakly on $\beta$ in the continuum limit.
The data of Fig.~\ref{fig:quenched} support this, with the proviso that even on
$32^3$ there is no sign of the saturation as $L\to\infty$ predicted by
(\ref{eq:polyrho}).
 
With a few more assumptions it is possible to estimate $\Delta m$. The density
of $m$ ($\bar m$) is proportional to $\xi e^{\pm gV(r)}$, where 
$g=2\pi\surd\beta$ is given by Dirac quantisation. With the (surely too crude)
approximation that $m$ and $\bar m$ densities are uncorrelated we can write
\be
\Delta m\propto\xi\int d^3r[\cosh(2\pi\surd\beta V(r))-1].
\ee
For large $r$ the integrand may be approximated as $2\pi^2\beta V^2$, so
\be
\Delta m
\propto \surd\beta e^{-C\beta/2}
\exp(-4\pi\surd\beta e^{-C\beta/2})\;\;\Rightarrow\;\;
{{\Delta
m}\over\rho}\propto\surd\beta e^{-C\beta/2}.
\label{eq:scaling}
\ee
The right panel of Fig.~\ref{fig:quenched} shows that the ratio $\Delta m/\rho$
predicted in (\ref{eq:scaling}) is qualitatively correct for
$\beta^\prime$\gesim1,
and moreover volume independent as $L\to\infty$, suggesting that the simulation
is able to probe the dilute plasma characteristic of the continuum limit. 
Note that the $\beta$-axis has
been reparametrised,\cite{Stack}
\be
\beta^\prime(\beta)
=\left[2\ln\left({{I_0(\beta)}\over{I_1(\beta)}}\right)\right]^{-1},
\ee
to match the Wilson to the Villain form of the lattice QED$_3$ action, for
which the Polyakov picture should be more accurate. Unfortunately 
quantitative agreement is less good; our
data yield a value for $C$ smaller than the analytic 
prediction,\cite{BMK,Ambjorn}
and the positive curvature at large $\beta$ is hard to explain, probably
requiring a better treatment of $m$ - $\bar m$ correlations.

\section{Results}

We now present results obtained from $O(10^6)$ HMD simulation
trajectories of each of the actions (\ref{eq:action}) and
(\ref{eq:maction},\ref{eq:fmaction}) on 
$16^3$, $20^3$ and $24^3$ lattices, using both $\beta_s=2.0$ and $\beta_s=4.0$.

\begin{figure}
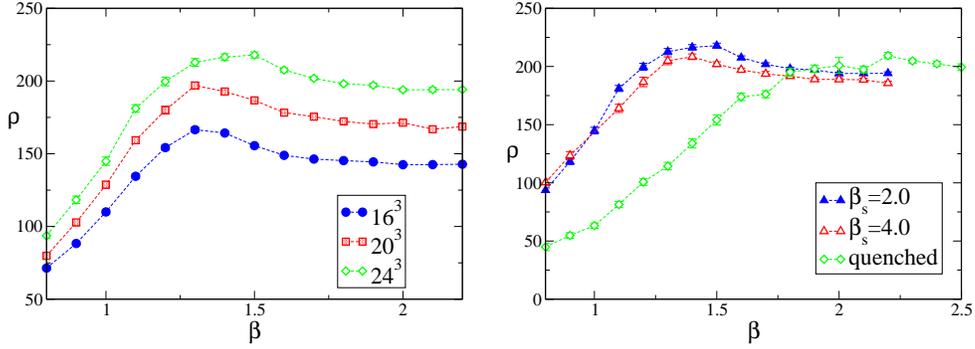

\begin{center}
\psfig{figure=rho_bs20.eps,height=1.8in}\hskip0.5cm
\psfig{figure=rho_L24.eps,height=1.8in}
\caption{$\rho$
vs. $\beta$ for $\beta_s=2.0$ on various lattices (left), and for different
$\beta_s$ on $24^3$ (right)
\label{fig:rho}}
\end{center}
\end{figure}
The left panel of Fig.~\ref{fig:rho} shows $\rho(\beta)$ for the three
different volumes at $\beta_s=2.0$. The curves peak at $\beta\approx1.3$,
with no significant evidence for the position of the peak changing with volume.
As in the quenched case,
$\rho$ increases with volume and there is no sign of saturation; however, there
is also no sign of a divergence, either at a particular $\beta^*$,
signalling a phase transition, or for a range of values of $\beta$, signalling
an infinite free energy for the $q\not=0$ sector and hence a phase in which
magnetic charge is confined. In the right hand panel we compare the 
$24^3$ results
of the two fermionic models with the corresponding quenched data. 
There is remarkably little
difference between $\beta_s=2.0$ and $\beta_s=4.0$; the main effect seems to be
a shift of the peak towards stronger coupling compared with quenched, 
a generic result of electric
charge screening due to virtual fermion pairs. The peak location is independent
of $\beta_s$, unlike the apparent location of the chiral transition $\beta_c$
seen in Fig.~\ref{fig:overview}.
So far there is no reason to interpret the $\rho(\beta)$ data as describing
anything other than a crossover from strong to weak coupling, as in the quenched
case.

\begin{figure}
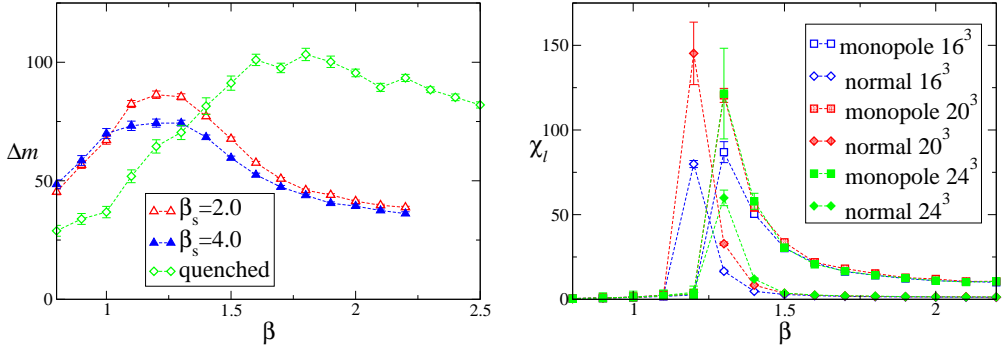

\begin{center}
\psfig{figure=deltam_L24.eps,height=1.8in}\hskip0.5cm
\psfig{figure=chi_l_bs20.eps,height=1.8in}
\caption{$\Delta m$ vs. $\beta$ for various $\beta_s$ on $24^3$ (left), 
and $\chi_l$ vs. $\beta$ for $q=0,1$ on various lattices (right)
\label{fig:deltam_chi}}
\end{center}
\end{figure}
The left hand panel of Fig.~\ref{fig:deltam_chi} tells a similar story for 
$\Delta m$. 
This quantity probes the cloud of virtual $m\bar m$ pairs produced
by the plasma to screen 
the external magnetic charge.
Once again there is a well-defined peak, this time at
a slightly stronger coupling, and still well to the left of the quenched peak.
An interesting feature is that the two fermionic curves only differ
significantly in shape in the range 1.0\lesim$\beta$\lesim 1.4 --
Fig.~\ref{fig:overview} confirms that in this region the fermions
are massive for the $\beta_s=2.0$ theory but massless (or at least much lighter)
for $\beta_s=4.0$. This suggests that light fermions are able to suppress the 
production of virtual $m\bar m$ pairs to some extent. There is, however, no
evidence that the large-$\beta$ behaviour differs from the quenched theory.

Finally, the right hand panel of  Fig.~\ref{fig:deltam_chi} shows the chiral 
susceptibility $\chi_l=\partial\langle\bar\psi\psi\rangle/\partial{\cal M}$ (or
at least its disconnected component) versus $\beta$, separately in both 
``normal'' $q=0$ and ``monopole'' $q=1$ vacua on the three different volumes.
The peaks are consistent with the chiral phase transition at $\beta_c\simeq1.4$
seen in Fig.~\ref{fig:overview}, but it is interesting that chiral symmetry
restoration seems to occur at a slightly weaker coupling in the presence of an 
external monopole. This can be understood semi-classically; the energy of a 
spin-singlet $\psi\bar\psi$ pair is lowered in the presence of a magnetic field
since the constituents have parallel magnetic moments, so pair condensation is
promoted. The vacuum therefore contains more $\psi\bar\psi$ pairs in the
vicinity of the external monopole, and the resulting inhomogeneity in
$\langle\bar\psi\psi\rangle$ increases $\chi_l$. 
Fig.~\ref{fig:deltam_chi} shows, however, that at least at large $\beta$ there
is no significant increase of this effect with volume, suggesting that the
spatial region over which pair enhancement takes place is finite in extent, 
implying in turn that the external monopole's field is still Debye-screened
by the $m\bar m$ plasma even in the presence of light dynamical fermions.

In summary, the data so far are consistent with the scenario that the monopole
plasma survives the introduction of dynamical fermions, even once chiral
symmetry is apparently restored. We have identified a region of crossover from
strong to weak-coupling behaviour similar in nature to that seen in the quenched
theory, and there is no sign of any phase transition or singular free energy in
the magnetic sector. Moreover there is 
no obvious relation between the chiral transition and the crossover.
At this stage, therefore, we favour the scenario of Ref.~10 over that of Ref.~9,
with all the usual caveats about the need for a more quantitative data analysis,
better data in the critical regions, and better understanding of the volume
scaling\dots

While we have at least a provisional answer to the question of monopole
survival,
the nature of the continuum limit of compact QED$_3$ remains
intriguing. 
Is chiral symmetry truly restored, at least for $N_f\geq4$? If so, we
have an example of a theory which is both ``confining'' and chirally symmetric! 
Perhaps this can be understood 
in the sense of a finite-ranged interaction between conserved
fermion currents; we can speculate whether the continuum limit coincides with
those
of the $\chi U\phi_3$ fermion-gauge-scalar model,\cite{chiUphi} or even the
$3d$ Thirring model~\cite{HL} at their UV fixed points.
An alternative scenario~\cite{Herbut} is that chiral symmetry 
remains broken in the continuum limit, but at a scale far too small to be
detected on currently available lattices. In this case the 
distinction between compact and
non-compact theories persists
as the issue of whether the photon is gapped or not.
What is clear is that QED$_3$, apparently
the simplest of gauge theory models,
continues to tantalise us even after more than 15 years of numerical study.

\section*{Acknowledgments}
SJH was supported in part by a PPARC Senior Research Fellowship, 
JBK by NSF grant PHY 03-04252, and BL 
by a Royal Society University Research Fellowship.
The bulk of the
simulations were performed on a dual-opteron Beowulf cluster at ETH
Z\"urich.

\end{document}